\documentclass[aps,prl,twocolumn,groupedaddress,showpacs]{revtex4}
\usepackage{graphicx}
\newcommand{\delete}[1]{}

\newcommand{\be}{\begin{equation}}
\newcommand{\ee}{\end{equation}}

\def\beq{\begin{equation}}
\def\eeq{\end{equation}}
\def\bea{\begin{eqnarray}}
\def\eea{\end{eqnarray}}
\def\ba{\begin{array}}
\def\ea{\end{array}}

\begin{document}

\title{Detecting high-frequency gravitational waves with optically-levitated sensors}
\author{Asimina Arvanitaki}
%\email[]{aarvan@stanford.edu}
\affiliation{Department of Physics, Stanford University, Stanford, CA 94305}
\author{Andrew A. Geraci}
%\email[]{ageraci@unr.edu}
\affiliation{Department of Physics, University of Nevada, Reno, NV 89557}

\date{\today}
\begin{abstract}

We propose a tunable resonant sensor to detect gravitational waves in the frequency range of 50 -- 300 kHz using optically trapped and cooled dielectric microspheres or micro-discs. The technique we describe can exceed the sensitivity of laser-based gravitational wave observatories in this frequency range, using an instrument of only a few percent of their size. Such a device extends the search volume for gravitational wave sources above 100 kHz by 1 to 3 orders of magnitude, and could detect monochromatic gravitational radiation from the annihilation of QCD axions in the cloud they form around stellar mass black holes within our galaxy due to the superradiance effect.
\end{abstract}


\maketitle

{\it{Introduction.}} Over the past 40 years optical trapping of dielectric objects, both macroscopic and atomic, has made a profound impact in a wide range of fields ranging from fundamental physics to the life sciences.  First studied by Ashkin and coworkers \cite{ashkin1}, optically trapped dielectrics in ultra-high vacuum become well decoupled from their room temperature environment \cite{omori,aerosols,aerosol3,kimble,cirac}. Recent work suggests that the center of mass motion of such levitated objects can attain mechanical quality factors
in excess of $10^{12}$, while internal vibrational modes are completely decoupled.  This remarkable decoupling can be harnessed for cooling the center of mass motion of such objects to the quantum ground state \cite{kimble,cirac,raizen,rochester}.  These systems also have been considered in the context of reaching and exceeding the standard quantum limit of position measurement \cite{libbrecht}. In addition, these techniques enable ultra-sensitive force detection \cite{rugar2,yocto,shortrange} and extend the study of quantum coherence to the mesoscopic regime.

In this paper, we study how nano- and micro-scale sensors trapped inside a medium-finesse optical cavity can be used to detect high frequency gravitational wave (GW) radiation. While there has been convincing indirect evidence for the existence of GWs \cite{hulsetaylor}, their direct detection has yet to be demonstrated. Such a detection is highly likely in the next decade with the new generation of laser-interferometer gravitational wave observatories \cite{ligo,advligo,virgo,geo,LCGT}, and will launch the field of gravitational wave astronomy.  While these detectors have been optimized in the frequency band of $10-10^4$ Hz, their sensitivity decreases at higher frequency due to photon shot noise.

We propose an alternative form of detector for improved sensitivity in the frequency range of $50-300$ kHz that does not rely on a shot-noise limited displacement measurement of test mass mirrors, but rather depends on a precision force measurement on the resonant harmonically trapped sensor. The detector we describe can yield sensitivities improved by more than an order of magnitude in this frequency band when compared with existing interferometers, while being only a fraction of their size.  The approach extends the effective search volume for sources between 100 and 300 kHz by $\sim 10 - 10^3$ when compared with Advanced LIGO \cite{advligo}.

Finally, we discuss GW sources at high frequencies. We focus on GW signals from the effects of the QCD axion on stellar mass black holes (BHs) through BH superradiance \cite{BH}. This novel signal comes from axion annihilation to gravitons, is monochromatic, long-lived and extremely different from all known astrophysical sources.

\begin{figure}[!t]
\begin{center}
\includegraphics[width=0.8 \columnwidth]{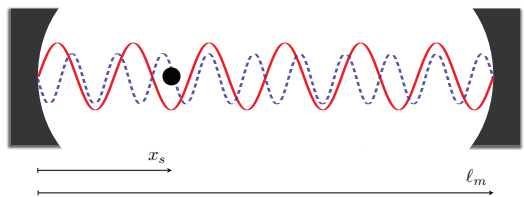}
\caption{A dielectic nanosphere or microdisc is optically trapped in an anti-node (solid red) of a cavity of length $\ell_m$ at position $x_s$. A second light field with two different frequency components (dashed blue) is used to cool and read out the axial position of the levitated object, respectively. Two additional beams perpendicular to the cavity axis (not shown) are used to cool the transverse motion of the sensor. A passing gravitational wave at frequency $\omega_{gw}$ displaces the sensor from its equilibrium position in the optical trap and imparts a force as described in text. The resulting displacement is resonantly enhanced when $\omega_{gw}$ coincides with the trap frequency $\omega_0$.}
\label{Fig:Experiment}
\end{center}
\end{figure}

{\it{Experimental Setup.}} The apparatus is schematically shown in Fig. \ref{Fig:Experiment}. We consider a dielectric sphere or microdisc with dielectric constant $\epsilon$, mass $m$, and density $\rho$ optically trapped and cooled in a cavity of length $\ell_{m}$ using two light fields of wavevector $k_t=2\pi/\lambda_{\rm{trap}}$ and $k_c=2\pi/\lambda_{\rm{cool}}$, respectively. The sphere or disc, which we generally refer to as the sensor in the following discussion, is levitated in an anti-node of the trapping light located near the input mirror ({\it{e.g.}} within $\ell_{m}/100$). We assume $\lambda_{\rm{trap}}$ is fixed and the cavity length is actively stabilized at low frequencies e.g. $<10$ kHz.

In the locally Lorentz (LL) frame with origin at the input mirror, a high-frequency gravitational wave, $h\equiv h(t-\frac{y}{c})$, that propagates in a direction perpendicular to the cavity axis will change the proper distance between the mirrors as well as the distance between the mirror and the dielectric sphere or disc (see Fig. \ref{Fig:Experiment}), respectively:
\bea
L_{m}=(1+\frac{1}{2} h) \ell_{m},\\
X_{s}=(1+\frac{1}{2} h) x_{s}.
\eea
With $\ell_{m}$ and $x_{s}$ we denote the proper distances between the mirrors, and between the mirror and the sensor in the absence of the GW. For the GW frequencies we are considering, corrections of order $h \frac{x^2}{\lambda_{GW}^2}$ can be safely neglected.  The electric field component of the trapping light with frequency $\omega=k_t c$ in the cavity is the sum of two counter-propagating waves
\be
E(t,x)=E_o (e^{-i \omega t + k_t X}-e^{-i \omega t - k_t X+2 k_t L_{m}}).
\ee
In the absence of a GW, the sensor is located in an antinode of the trapping field at position $x_{s} = x_{\rm{min}}$. In contrast to the sphere position $x_s$, the change in $x_{\rm{min}}$ in the presence of the GW is not determined by the proper distance change between two massive particles but by the interference of the two counter-propagating waves in the cavity. The condition for the antinode in the absence of the GW is given by $k_t (\ell_m-x_{\rm{min}}) =(n+1/2)\pi$ for an integer $n$.
By requiring the same condition is satisfied under the influence of the GW, we obtain the new position of the trap minimum at the position of the sensor: $k_t (L_m-x'_{\rm{min}})=k_t (\ell_m-x_{\rm{min}})$. The shift of the trap minimum is $\delta X_{\rm min}=\frac{1}{2}h \ell_m$, while the shift of the sensor position is $\delta X_s= \frac{1}{2}h x_s$. The difference of these quantities gives us the displacement of the sensor with respect to its trap minimum under the influence of the GW:
\be
\label{eq:gweffect}
\Delta X =\delta X_s-\delta X_{\rm{min}}= \frac{1}{2}h (x_{s}-\ell_{m})+ {\mathcal{O}}(h^2)
\ee
The above equation shows that the sensor needs to be placed close to the input mirror, in order to maximize the effect of the gravitational wave, while the effect is zero at the position of the end mirror.

Treating the GW as coming from a monochromatic source of frequency $\omega_{gw}$ and amplitude $h_o$, there is a oscillatory driving force on the sphere:
\be
F_{gw}= - \frac{m \omega_{gw}^2}{2} (x_{s}-\ell_{m}) h_o \cos (\omega_{gw} t +\Delta \phi)
\label{signal}
\ee
When the trapping frequency of the sphere matches $\omega_{gw}$, the sphere will be resonantly excited.  In this way the device operates like a resonant-bar detector, except with a wide tunability \cite{bardetector}.

For concreteness, we consider a
cavity of length $\ell_m=100$ m, finesse ${\mathcal{F}}=10$, and cavity mode waist
$w=75$ $\mu$m. The cavity has g-parameters $g_1 \approx -985$ and $g_2 \approx -1\times 10^{-3}$. In order to avoid significant beam clipping, we assume the input mirror has a radius of 1 cm, while the end mirror has an (active) radius of 1 m. We note that this mirror is somewhat larger than those previously used for cavities and would likely require custom fabrication tools. %Alternatively the beam size at the end mirror could be reduced by adding lenses into the system, enabling longer cavities.
We consider a silica ($\epsilon=2,\rho=2.3$ kg/m$^3$) nanosphere with radius $a = 150$ nm or an apodized microdisc with radius $a = 75$ $\mu$m and thickness $t=0.5$ $\mu$m, respectively. In Table \ref{table1}, expressions for the trapping frequency $\omega_0$, trap depth $U$, gas-damping coefficient $\gamma_g$, and scattered photon-recoil heating rate $\gamma_{sc}$ are given in terms of the average intensity over the disc or sphere surface $I_t$, the background gas pressure $P$ and mean speed $\bar{v}$,
the volume of the microdisc or sphere $V$, and the cavity mode volume $V_c=\pi w^2 L /4$, following Refs. \cite{kimble,kimblemem}. Here ${\mathcal{F}}_{disc}$ is the micro-disc limited cavity finesse \cite{kimblemem} and the optomechanical coupling of the cooling mode is $
g$.

\begin{table}[!t]
\begin{center}
  \begin{tabular}{@{}ccc@{}}
  \hline
  \hline
  Quantity & nanosphere & microdisc \\
  \hline
$\omega_0$ & $\left[ \frac{6k_t^2I_t}{\rho c}
{\mathcal{R}}e \frac{\epsilon-1}{\epsilon+2} \right]^{1/2}$ & $\left[ \frac{2k_t^2I_t}{\rho c}
{\mathcal{R}}e (\epsilon-1)\right]^{1/2}$ \\
$U$ & $\frac{3I_{t}V}{c} \frac{\epsilon-1}{\epsilon+2}$ & $\frac{I_{t}V}{c} ({\epsilon-1})$ \\
$g$ &  $\frac{3V}{4V_c} \frac{\epsilon-1}{\epsilon+2} \omega_c$ & $\frac{V}{4V_c} (\epsilon-1) \omega_c$ \\
$\gamma_g$ & $\frac{16 P}{(\pi \bar{v}
\rho a)}$ & $\frac{32 P}{(\pi \bar{v}
\rho t)}$\\
$ \gamma_{sc} $ & $\frac{2}{5} \frac{\pi^2 \omega_0 V}{\lambda^3} \frac{(\epsilon-1)}{(\epsilon+2)}$ & $\frac{V_c}{V}\frac{\lambda}{4L}\frac{1}{(\epsilon-1){\mathcal{F}}_{\rm{disc}}}\omega_0$\\
  \hline
  \hline
  \end{tabular}
\caption{\label{table1} Expression for trapping and cooling parameters.}
\end{center}
\end{table}

The cavity is driven with a trapping laser of wavelength
$\lambda_{\rm{trap}}=1.55$ $\mu$m and power $P_t=7.7$ W or $P_t=9.6$ W for a sphere and disc, respectively, corresponding to an axial trap frequency of $\omega_0/2\pi = 100$ kHz. The
cooling light has wavelength $\sim 1.55$ $\mu$m, frequency $\omega_c$, input power $P_c=1.1$ W or $2.2$ $\mu$W for the sphere or disc, respectively, and an optimized red
detuning of $\delta$. The cavity decay rate is
$\kappa = \pi \nu_0 / {\mathcal{F}}$, and $\nu_0$ is the
free-spectral range. Additional cavity loss due to photon scattering is negligible: less than $10^{-3} \kappa$ for our parameters. The microsphere or disc absorbs optical power from both the trapping and
cooling light in the cavity, which results in an increased internal
temperature $T_{\rm{int}}$ \cite{fusedsilicaloss}.  Assuming negligible cooling due to gas
collisions, the absorbed power is re-radiated as blackbody
radiation.
%$T_{\rm{int}}$ is listed in Table \ref{table2} for silica \cite{fusedsilicaloss} %with
%dielectric response $\epsilon=\epsilon_1+i\epsilon_2$, with
%$\epsilon_1=2$ and $\epsilon_2=1.0 \times 10^{-7}$ as in Ref.
%\cite{fusedsilicaloss}, and $\epsilon_{\rm{bb}}=0.1$ as in Ref.
%\cite{kimble},
%for an environmental temperature $T_{\rm{ext}}=300$K.
$T_{\rm{int}}$ and $T_{\rm{CM}}$ are not significantly
coupled over the time scale of the experimental measurements at
$P_{\rm{gas}}=10^{-11}$ Torr. Other experimental parameters are shown in Table \ref{table2}.

The Gaussian profile of the trapping beam near the
mode waist provides transverse confinement, with an oscillation
frequency of $\sim 320$ Hz. %The earth's gravitational field displaces the equilibrium position of the sphere or disc by $\sim 2.2$ $\mu$m.
Transverse motion can be cooled with active feedback by
modulating the power of transverse lasers using the signal
from a transverse position measurement.%, for example generated by
%measuring scattered light incident on a quadrant photodiode.
We assume a modest cooling factor of $\lesssim 100$ in the transverse directions
to counteract the effects of recoil heating and localize the sensor.

For detecting the axial position of the sensor, light from the cavity-cooling laser can be frequency shifted to be on-resonance with the cavity to maximize sensitivity. The phase of the detection light reflected
from the cavity is modulated by the sensors motion through the
optomechanical coupling $\partial{\omega_c}/\partial{z}=k_c g$.
Photon shot-noise limits the minimum detectable phase shift to
$\delta \phi \approx 1/(2\sqrt{I})$ where $I \equiv P_d /(\hbar
\omega_c)$ \cite{hadjar}. The corresponding photon shot-noise
limited displacement sensitivity
 is $ \sqrt{S_z(\omega)}= \frac{\kappa}{4k_cg}
\frac{1}{\sqrt{I}}\sqrt{1+\frac{4\omega^2}{\kappa^2}}$
\cite{kippenbergdisp}, for an impedance matched cavity.  In our case, a detection power $P_d = .2$ mW or $1.1$ W for a disc or sphere, respectively, corresponds to  $ \sqrt{S_z(\omega)} = 3.8 \times 10^{-16}$ m/$\sqrt{{\rm{Hz}}}$, and $2.7 \times 10^{-12}$ m/$\sqrt{{\rm{Hz}}}$. The thermally-driven resonant CM motion of the sensor is typically much greater e.g. $10^{-14}$m/$\sqrt{{\rm{Hz}}}$ for a microdisc, and this CM thermal motion sets the sensitivity limit for the experiment. Beating the standard quantum limit is not required over the frequency band of interest for either the disc or sphere sensor. Surface motion due to internal thermoelastic and Brownian thermal noise remains more than one order of magnitude below the resonant CM thermal motion, taking a silica disc loss factor of $10^{-5}.$ The effects of internal thermal motion are further suppressed since the disc or sphere acts like a refractive (rather than reflective) element in the cavity: the output displacement signal depends on the CM motion.  We assume that substrate vibrational noise, electronics noise and laser noise can be controlled at a level comparable to the photon shot noise.

\begin{figure}[!t]
\begin{center}
\includegraphics[width=1.0\columnwidth]{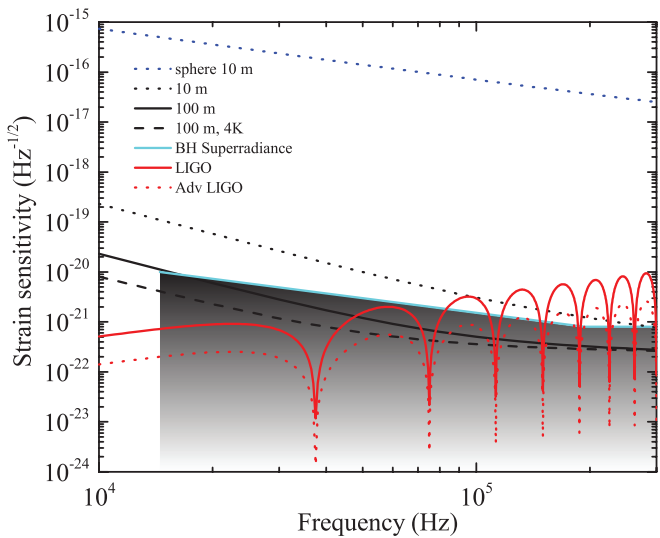}
\caption{(color online) Strain sensitivity for optically levitated micro-discs (black) or spheres (blue) for experimental parameters described in the text. For comparison, also shown are the LIGO and predicted Advanced LIGO sensitivity in the frequency range of 10-300 kHz \cite{ligocurves,malik}.  The shaded region denotes predicted signals due to Black Hole superradiance.
\label{strain}}
\end{center}
\end{figure}

The cooling serves to damp the $Q$ factor to $Q_{\rm{eff}}$ so that perturbations to the system
ring down within reasonably short periods of time, to reduce the requirement on the laser intensity stabilization, and to mitigate heating due to the recoil of trap laser photons. At the same time the mode temperature is reduced to $T_{\rm{eff}}$.  The minimum detectable force due to thermal noise at temperature
$T_{\rm{eff}}$ is $F_{\rm{min}} = \sqrt{\frac{4 k k_B T_{\rm{eff}}
b}{\omega_0 Q_{\rm{eff}}}},$
where $k$ is the center-of-mass mode spring
constant, and $b$ is the measurement bandwidth. The thermal-noise limited minimum detectable strain due to a GW will be approximately
\be
h_{\rm{limit}} = \frac{4}{\omega_0^2 \ell_m} \sqrt{\frac{k_B T_{\rm{eff}} \gamma_g b }{ m} \left[1+\frac{\gamma_{\rm{sc}}+R_{+}}{n_i \gamma_g}\right]} H(\omega_0)
\label{heq}\ee
where the cavity response function $H(\omega_0) = \sqrt{1+(2{\mathcal{F}}/\pi)^2 \sin^2{(\omega_0 \ell_m /c)}}.$
We define a factor $\chi = \frac{\gamma_{\rm{sc}}+R_{+}}{n_i \gamma_g}$ which describes the importance of photon recoil heating $\gamma_{\rm{sc}}$ and the efficiency of the cavity cooling.  The factor $R_{+}$, defined in Ref. \cite{kimble} can be minimized by going into the resolved sideband regime and can be generally neglected when compared with $\gamma_{sc}$.  There are two general regimes of scaling, $\chi << 1$ and $\chi >>1$.  For $\chi <<1$, the effects of photon recoil do not significantly degrade the force sensitivity, and for a nanosphere, $h_{\rm{lim}} \propto \omega_0^2 r^2 T^{-1/4} P^{-1/4}$.  In the regime $\chi >>1$, photon recoil heating becomes significant, and damping without an equal amount of cooling occurs.  Here for a nanosphere the sensitivity scales as $h_{\rm{lim}} \propto \omega_0$ and is independent of $r,T$ and $P$.   The micro-disc geometry scatters much less light, as pointed out in Ref \cite{kimblemem} and recoil heating is significantly reduced. We assume a disc-limited cavity finesse of $10^5$, which is reasonable for an apodized disc \cite{kimblemem}.  In Fig. \ref{strain} we plot the expected gravitational wave strain sensitivity using the selected experimental parameters.  The sphere sensor operates in the regime $\chi>>1$, while the disc transitions between $\chi<1$ and $\chi>1$ as frequency increases, resulting in a different slope.  Due to reduced light scattering and larger mass, the disc has superior sensitivity. We include the sensitivity of current and near-term future detectors such as LIGO and Advanced LIGO. The LIGO sensitivity continues to decrease at frequencies above its free-spectral range, as the wavelength of the corresponding GW becomes shorter than the instrument size, reducing the GW transfer function \cite{malik}.

\begin{table}[!t]
\begin{center}
  \begin{tabular}{@{}cccc@{}}
  \hline
  \hline
  Parameter & Units & nanosphere & microdisc \\
  \hline
%$r$ & nm & 200 & 1500 \\
$\lambda_{\rm{trap}}$ & $\mu$m & $1.55$ & $1.55$ \\
%$P$ & W & $0.035-3.5$ & $1.2$ \\
%$w$ & $\mu$m & $25$ & $250$ \\
%$U/k_B$ & K & $2.56 \times 10^4 $ \\
$\omega_0/2\pi$ & Hz & $1 \times 10^5$ & $1 \times 10^5$ \\
$a$ & $\mu$m & $0.15$ & $75$\\
$w_0$ & - & $75$ & $75$ \\
$ T_{\rm{int}} $ & K & $547$ & $743$\\
$T$ & K & $300$ & $300$ \\
%$\delta z$ & $\mu m$ & $2.2$ \\
$ Q,(Q_{\rm{eff}}) $ & - & $1.5 \times 10^{13},(3.4 \times 10^{6})$ & $2.5 \times 10^{13},(5.4 \times 10^{5})$\\
%\hline
$\delta/\kappa$ & - & $-.22$ & $-.50$\\
%$\zeta$ & - & $0.000588 $ & $0.002$\\
%$ g_2/2\pi $ & Hz & $4.1 \times 10^4 $ & $2.4 \times 10^4 $\\
%$\kappa_{\rm{sc}}/\kappa $ & - & $0.27$ & $0.19$ \\
$ n_T,(n_f) $ & - & $6.3 \times 10^7$,$(1.3 \times 10^4)$ & $6.3 \times 10^7$,$(2.6)$\\
%\hline
%$ \sqrt{S_z} $ & m$/\sqrt{\rm{Hz}}$ & $3.8 \times 10^{-14}$ \\
$h_{\rm{min}}$ & 1$/\sqrt{\rm{Hz}}$ & $7.0 \times 10^{-17}$ & $5.0 \times 10^{-22} $ \\
$\sqrt{1+\chi}$ & $-$ & $31$ & $1.39$\\
%$ z_{\rm{th}} $ & m$/\sqrt{\rm{Hz}}$ & $3.8 \times 10^{-11} $ \\
  \hline
  \hline
  \end{tabular}
\caption{\label{table2} Experimental parameters for trapping and
cooling a silica sphere with radius $a=150$ nm in a $10$ m cavity and for a microdisc in a $100$ m cavity.}
\end{center}
\end{table}

We expect a passive vibration isolation system similar to that employed in LIGO would be sufficient for the proposed measurements, particularly since the effects of mechanical vibration become less significant at the high operating frequencies.  To properly distinguish GWs from other disturbances, two perpendicular arms of an interferometer can be used, as in existing GW observatories.  Also, coincidence between several operating sensors can be used to discriminate other backgrounds.  Motion in the end mirrors or their coatings due to thermal noise are not significant at the high operating frequencies.
%Low frequency fluctuations or drift in the cavity length or laser frequency can modulate the intra-cavity intensity and therefore cause fluctuations $\delta \omega_0$ in the trapping frequency. We require $\delta \omega_0$ $<<$ $\omega_0/Q_{\rm{eff}}$, which can be accomplished with active stabilization.

For a sphere, additional motion ({\it{e.g.}} due to rotation) generally occurs at different frequencies and can be averaged out in a measurement. For a microdisc, although the restoring force of the laser tends to keep the disc upright, there can be torsional modes and axial rotation.  The frequencies of these modes can be separated from the center of mass mode. Also, flexural modes within the disc can be neglected if their frequency is large compared to the variation of the effective trapping frequency due to laser intensity variation over the disc surface.  We can therefore treat the disc as a rigid object for the proposed parameters, as the fundamental mode (2,0) of its deformation has a frequency of 126 kHz. % , which is larger than the variation in trapping frequency over the extent of the disc for $\omega_0/(2\pi) = 10^5$ Hz. %This variation can be further reduced by increasing the beam waist to $1.5 a$, which still allows ${\mathcal{F}}_{\rm{disc}} \sim 10^5$ \cite{kimblemem}.  This approximation becomes less accurate at higher trapping frequencies.

{\it{GW sources}.}  The most interesting source of GWs in the high frequency regime arises from the effect of the QCD axion on astrophysical stellar mass BHs through the Penrose superradiance process \cite{BH}. The QCD axion is a pseudo-Goldstone boson that naturally solves the strong CP problem and explains the smallness of the neutron's electric dipole moment \cite{axion}. Non-perturbative QCD effects generate a cosine potential that gives the axion a mass: $\mu_a=6 \times 10^{-10} eV/c^2 \left(10^{16}~\text{GeV}/f_a\right),$
where $f_a$ is the axion decay constant.

The Compton wavelength of the QCD axion with $f_a\gtrsim 10^{16}~\text{GeV}$ matches the size of stellar mass BHs and allows for the axion to bind with the BH ``nucleus" forming a gravitational atom in the sky. The occupation number of the levels that satisfy the superradiance condition grows exponentially extracting energy and angular momentum from the BH. At the end of the process there is a Bose Einstein axion condensate cloud surrounding the BH. Axions in this cloud produce GWs through transitions between different atomic levels and through annihilations of axions to one graviton. The latter process becomes possible since the axion is its own antiparticle and the BH ``nucleus" makes sure that energy and momentum are conserved, exactly as a positron can annihilate with an atomic electron to a single photon. For annihilations, the frequency of the produced GWs is given by twice the mass of the axion: $f=145~\text{kHz}  \left( 2\times 10^{16}~\text{GeV}/{f_a}\right)$ which lies in the optimal sensitivity range of our setup when $f_a$ is between $10^{16}$ and $2\times 10^{17}$ GeV. The signal is coherent, monochromatic and thus completely different from all ordinary astrophysical sources.

The GW amplitude coming from annihilations is:
\be
h\sim 10^{-19} \left(\frac{\alpha}{\ell}\right)\epsilon \left(\frac{10~\text{kpc}}{r}\right) \left(\frac{M_{BH}}{2\times M_{\tiny{\odot}}}\right)
\label{annsignal}
\ee
where $\alpha=G_N M_{BH} \mu_a$ and $\ell$ is the orbital quantum number of the super radiating level \cite{BH}. The ratio $\frac{\alpha}{\ell}$ is constrained by the superradiance condition and it can acquire a maximum value of $\approx 0.5$. Finally, $\epsilon$ is the fraction of the BH mass the axion cloud carries and it can be as high as $10^{-3}$. In Fig. \ref{strain}, we compare the annihilation signal for a source 10 kpc away to the experimental sensitivity assuming $10^6$ sec of integration time. The slope of the curve is determined by the maximum BH mass that can superradiate for the given axion mass $-$ it decreases linearly with increasing axion mass. The curve saturates when this maximum BH mass equals the smallest possible BH that can be formed through astrophysical processes, $\approx 2 M_{\odot}$. For a GUT scale $f_a$ axion, a signal coming from within our galaxy could be detected. The number of BHs within our galaxy in estimated to be $10^7-10^9$ and the axion annihilation signal can last for a few weeks or up to a year, so it is likely that the proposed setup has a good chance of detecting it.

Astrophysical gravity wave sources have a natural upper bound on the GW frequency they can produce. This is determined by the lower bound on the black hole mass produced through ordinary stellar dynamics: $f_{max}\approx c^3/G_{N} M_{BH_{min}},$
where $G_N$ is Newtons constant and $M_{BH_{min}}$ is the minimum black hole mass. This places $f_{max} \sim 30$ kHz for a one solar mass BH which is on the edge of the optimal experimental sensitivity.

{\it{Discussion.}} The method we have described promises to be the most sensitive approach for detecting GWs in the frequency range over 100 kHz. Although few astrophysical sources are likely to exist at such high frequencies, there can be a variety of sources associated with the early Universe and Beyond the Standard Model.  Such a source is the well-motivated QCD axion.

%The minimal requirements on vibration isolation at high frequencies relax limitations on the geographic location of the experiment.
Due to the resonant GW detection, the strain sensitivity of our setup is only limited by the thermal motion of the sensor and not by the laser shot noise. % $-$ the requirements on displacement sensitivity are relaxed by over $10^3$ when compared with e.g. LIGO.
Several beads or discs can be simultaneously trapped in different anti-nodes of the cavity, each with a different trap frequency determined by the local beam waist size. Also, $Q_{\text{eff}}$, which determines the bandwidth at a given frequency, can be tuned to as small as $\sim 10^3$ without significant loss in sensitivity. Such an instrument can thus scan over a wide variety of frequencies at the same time, which is crucial for the search for QCD axion signals.  It also may be possible to further improve the sensitivity by using focusing optics to extend the cavity length.

Pushing the sensitivity limit in this uncharted high frequency region adds to the development of GW astronomy that will take place in the next decade with Advanced LIGO and atom interferometry {\cite{AGIS}}. GWs propagate unperturbed after they are created, allowing us to study the remote corners of our Universe, and will become an indispensable tool for astrophysics and cosmology.

We thank Savas Dimopoulos, Sergei Dubovsky, Nemanja Kaloper, and Jonathan Weinstein for discussions, and the referees for useful comments.


\begin{thebibliography}{99}

\bibitem{ashkin1} A. Ashkin, Phys. Rev. Lett. {\bf{24}}, 156 (1970), A. Ashkin and J. M. Dziedzic, Appl. Phys. Lett. {\bf{19}}, 283 (1971), A. Ashkin and J. M. Dziedzic, {\it{ibid.}} {\bf{28}}, 333 (1976).
\bibitem{omori} R. Omori, T. Kobayashi, and A. Suzuki, Opt. Lett. {\bf{22}},816 (1997).
\bibitem{aerosols} D. McGloin {\it{et. al.}}, Faraday Discuss. {\bf{137}}, 335 (2008).
\bibitem{aerosol3} L. Mitchum and J. P. Reid, Chem. Soc. Rev. {\bf{37}}, 756 (2008).
\bibitem{kimble}  D. E. Chang {\it{et. al.}}, Proc. Nat. Acad. Sci. {\bf{107}}, 1005 (2010).
\bibitem{cirac}  O. Romero-Isart, M. L. Juan, R. Quidant, J. I. Cirac, New J. Phys. {\bf{12}}, 033015 (2010).
\bibitem{raizen}
T. Li, S. Kheifets, and M.G. Raizen,  Nature Physics \textbf{7}, 527 (2011).
\bibitem{rochester}
J. Gieseler, B. Deutsch, R. Quidant {\it{et. al.}}, arXiv 0426662 (2012).
\bibitem{libbrecht} K. G. Libbrecht and E. D. Black, Phys. Lett. A
{\bf{321}}, 99 (2004).
\bibitem{rugar2}
D. Rugar {\it{et. al.}}, Nature {\bf{430}}, 329 (2004).
\bibitem{yocto}
R. Maiwald {\it{et. al.}} Nature Physics {\bf{5}} 551 (2009), M. Biercuk {\it{et. al.}}, arxiv:1004.0780 (2010).
\bibitem{shortrange}
A. A. Geraci, S.B. Papp, and J. Kitching, Phys. Rev. Lett. {\bf{105}} 101101 (2010).
\bibitem{hulsetaylor}
Taylor, J.H., Fowler, L.A. and Weisberg, J.M., Nature 277, 437 (1979).
\bibitem{ligo}
B.Abbott, {\it{et. al.}}, Rep. Prog. Phys. {\bf{72}}, 076901 (2009).
\bibitem{advligo}
G. M. Harry (for the LIGO Scientific Collaboration), Class. Quantum Grav. 27 084006 (2010).
\bibitem{virgo}
T. Accadia {\it{et. al.}}, Journal of Inst. 7, P030012 (2012), The Virgo Collaboration, note VIR-027A-09 (2009).
\bibitem{LCGT}
K. Kuroda (for the LCGT Collaboration), Class. Quantum Grav. 27, 084004 (2010).
\bibitem{geo}
H. Grote (for the LIGO Scientific Collaboration), Class. Quantum Grav. 27 084003 (2010), B. Willke {\it{et. al.}} {\it{ibid.}} 23, S207 (2006).
\bibitem{BH}
%\cite{Arvanitaki:2009fg}
  A.~Arvanitaki, S.~Dimopoulos, S.~Dubovsky, {\it{et. al.}},
  %``String Axiverse,''
  Phys.\ Rev.\ D {\bf 81}, 123530 (2010);
  %%CITATION = ARXIV:0905.4720;%%
  A. Arvanitaki and S. Dubovsky, Phys.Rev.\textbf{D 83}, 044026 (2011).
\bibitem{bardetector} A. de Waard, et al., Class. Quantum Grav. \textbf{20}: S143–S151 (2003), O. D. Aguiar et. al., {\it{ibid.}} \textbf{25} 114042 (2008)
\bibitem{kimblemem} D.E. Chang et. al., New. J. Phys. \textbf{14}, 045002 (2012).
\bibitem{fusedsilicaloss}
R. Kitamura, L. Pilon, and M. Jonasz, Appl. Opt. {\bf{46}}, 8118 (2007).
\bibitem{hadjar} Y. Hadjar {\it{et. al.}}, Europhys. Lett. {\bf{47}}, 545 (1999).
\bibitem{kippenbergdisp} G. Anetsberger {\it{et. al.}}, Nature Physics {\bf{5}}, 909 (2009).
%\bibitem{epstein} P. S. Epstein, Phys. Rev. {\bf{23}}, 710 (1924).
\bibitem{ligocurves}
http://ligo.caltech.edu/advLIGO/scripts/ref\_des.shtml, https://dcc.ligo.org/, LIGO-T0900499, LIGO-T0900288.
\bibitem{malik}
M. Rakhmanov, Phys. Rev. D. {\bf{71}}, 084003 (2005), LIGO-T050136-00-W.
\bibitem{axion}
%\cite{Peccei:1977hh}
  R.~D.~Peccei and H.~R.~Quinn,
  %``CP Conservation in the Presence of Instantons,''
  Phys.\ Rev.\ Lett.\  {\bf 38}, 1440 (1977);
  %\cite{Weinberg:1977ma}
  S.~Weinberg,
  %``A New Light Boson?,''
  Phys.\ Rev.\ Lett.\  {\bf 40}, 223 (1978);
  %\cite{Wilczek:1977pj}
  F.~Wilczek,
  %``Problem of Strong p and t Invariance in the Presence of Instantons,''
  Phys.\ Rev.\ Lett.\  {\bf 40}, 279 (1978).
  %\cite{Dimopoulos:2008sv}
\bibitem{AGIS}
  S.~Dimopoulos, P.~W.~Graham, J.~M.~Hogan, {\it{et. al.}},
  %``An Atomic Gravitational Wave Interferometric Sensor (AGIS),''
  Phys.\ Rev.\ D {\bf 78}, 122002 (2008).
  %%CITATION = ARXIV:0806.2125;%%

\end{thebibliography}
\end{document}